\documentclass[10pt,conference,letterpaper]{IEEEtran}
\IEEEoverridecommandlockouts
\usepackage{cite}
\usepackage{amsmath,amssymb,amsfonts}
\usepackage{algorithmic}
\usepackage{graphicx}
\usepackage{textcomp}
\usepackage{xcolor}
\usepackage{algorithm}
\usepackage{algorithmic}
\usepackage{pgfplots}
\usepackage{stfloats}
\usepackage{makecell}
\usepackage{multicol}
\usepackage{multirow}
\pgfplotsset{compat=1.18}

\def\BibTeX{{\rm B\kern-.05em{\sc i\kern-.025em b}\kern-.08em
    T\kern-.1667em\lower.7ex\hbox{E}\kern-.125emX}}
\begin{document}

\title{DGNNFlow: A Streaming Dataflow Architecture for Real-Time Edge-based Dynamic GNN Inference \\ in HL-LHC Trigger Systems}

\author{
\IEEEauthorblockN{
Davendra Maharaj\IEEEauthorrefmark{1},
Tu Pham\IEEEauthorrefmark{1},
Peter Meiring\IEEEauthorrefmark{2},
Kyungmin Park\IEEEauthorrefmark{2},
Sena Durgut\IEEEauthorrefmark{2},
Cong Hao\IEEEauthorrefmark{1},
Matteo Cremonesi\IEEEauthorrefmark{2}}
\IEEEauthorblockA{\IEEEauthorrefmark{1}Georgia Institute of Technology, Atlanta, GA 30332, USA}
\IEEEauthorblockA{\IEEEauthorrefmark{2}Carnegie Mellon University, Pittsburgh, PA 15213, USA}
}

\maketitle

\begin{abstract}
Dynamic GNN inference exhibits strong capability to model interactions over time, such as complex particle collision events in High Energy Physics (HEP) experiments at High Luminosity Large Hadron Collider (HL-LHC). With much larger scale of collision data captured in future HEP experiments to help unlocking physics discoveries and limitation in both offline compute capacity and storage, revamped trigger systems require FPGAs to run ultra-low-latency Machine Learning models with low power consumption for online filtering of useful events. Many state-of-the-art GNN accelerators relied on static graph structures, but this assumption breaks down in HL-LHC trigger systems and other edge-based dynamic GNN applications where edge embeddings can change in-place based on neighbor node embeddings during runtime. We propose DGNNFlow, a novel streaming dataflow architecture for real-time edge-based dynamic GNN inference applications (including but not limited to HL-LHC trigger systems) along with three key contributions. First, we introduce hardware enhancement for edge embedding dynamic computation. Second, we alleviate data dependencies in edge-based dynamic GNN dataflow with Node Embedding Broadcast. Third, we provide input dynamic graph construction for complete support of graphs without pre-defined edge embeddings. We deploy DGNNFlow using AMD Alveo U50 FPGA to evaluate performance at 200~MHz clock frequency. DGNNFlow achieved 2.59x - 4.36x and 1.30x - 2.14x speedup compared to NVIDIA RTX A6000 GPU (batch sizes 1 and 2) with 3.59x - 3.70x less power consumption, achieved 2.29x - 3.54x speedup with 1.93x - 2.12x less power consumption compared to Intel Xeon Gold 6226R CPU. Our implementation is available on GitHub\footnote{https://github.com/sharc-lab/DGNNFlow}.
\end{abstract}

\section{Introduction}
\subsection{GNN, Edge-based Dynamic GNN, and Acceleration}
Graph Neural Network (GNN) is important class of neural network models for non-Euclidean graph data structures. They are utilized in various applications, such as recommendation systems, physics simulation, and social network analysis. GNNs iteratively update node embeddings by aggregating information from corresponding neighbors to capture both local and global graph structures \cite{YANG2024106083}~\cite{10.1007/978-3-031-43415-0_39}. Edge-based Dynamic GNNs such as EdgeConv~\cite{wang2019dynamicgraphcnnlearning} also update edge embeddings over time based on corresponding neighbor node embeddings to capture relationships among nodes in graphs.

GNN inference poses unique computational challenges, such as irregular memory access patterns and data-dependent execution paths in aggregation, so specialized GNN accelerators have emerged to address these challenges by exploiting fine-grained parallelism, low-latency datapaths, and customization tailored to GNN workloads.

\subsection{HEP experiments and Hardware Accelerator Demand}
Next-generation High Energy Physics (HEP) experiments designed for HL-LHC~\cite{HLLHC} will generate collision data every 25~ns, yielding data rates in order of petabytes per second, far beyond capacity of current data storage systems. To select potentially useful events in real time, two-level trigger system is employed and completely re-designed for the start of HL-LHC data-taking in 2030. Hardware-based Level-1 Trigger (L1T)~\cite{L1TTDR} will process limited-granularity detector data to make accept-reject decision, reducing event rate from 40~MHz to 750~kHz with ultra-low latency of 12.5~µs. L1T requires time-efficient, low-power processing and high-quality trigger quantities for complex trigger decisions. Given L1T's critical role in controlling event rates and GNN's importance in complex pattern recognition for particle physics, there is a pressing need for hardware accelerators supporting dynamic edge embedding inference and flexible message passing.

\subsection{Related Work and Limitations}
Recent GNN hardware accelerators have attempted to overcome inefficiencies of conventional CPUs and GPUs in graph-based computations. However, many existing designs primarily focus on static graphs and regular message-passing model, limiting their applicability to execute edge-based dynamic GNNs. HyGCN~\cite{HyGCN} uses hybrid aggregation to group nodes with same degrees, restricts irregular memory accesses, but primarily targets static graphs with fixed connectivity. AWB-GCN~\cite{AWBGCN} adapts processing units based on node degrees for workload rebalancing and better resource utilization, but also assumes static graphs and fixed adjacency matrices. GraphACT~\cite{GraphACT} dynamically adjusts processing resources in accordance with workloads primarily for static graphs. DGNN-Booster~\cite{DGNN-Booster} transfers sequence of entire static graph snapshots from host to FPGA over time for dynamic GNN and does not natively support in-place updates of dynamic graphs' edge embeddings directly on FPGA. FlowGNN\cite{FlowGNN} provides reconfigurable dataflow framework for wide range of GNN models, but assumes statically provided edge embeddings and fixed graph connectivity before inference.

\subsection{Our Contributions}
To address above demand and limitations, we propose our following contributions:
\begin{itemize}
\item Develop FPGA-based streaming dataflow architecture that extends FlowGNN\cite{FlowGNN} to support real-time edge-based dynamic GNN inference applications, including but not limited to HL-LHC trigger system.
\item Introduce hardware enhancement to compute edge embeddings dynamically on FPGA without needing to transfer sequences of static graph snapshots from host to FPGA over time.
\item Alleviate data dependencies in edge-based dynamic GNN dataflow without need for irregular memory access.
\item Provide input dynamic graph construction for graphs without pre-defined edge embeddings.
\item Prototype DGNNFlow on an AMD Alveo U50 FPGA and evaluate on-board performance with dataset generated using the DELPHES framework~\cite{Delphes}.
\end{itemize}

\section{Case Study: Dynamic GNN Algorithm \\ for HL-LHC Trigger Systems}
\subsection{L1DeepMETv2 Model Architecture}
\begin{figure}[b]
\centerline{\includegraphics[width=0.9\columnwidth]
{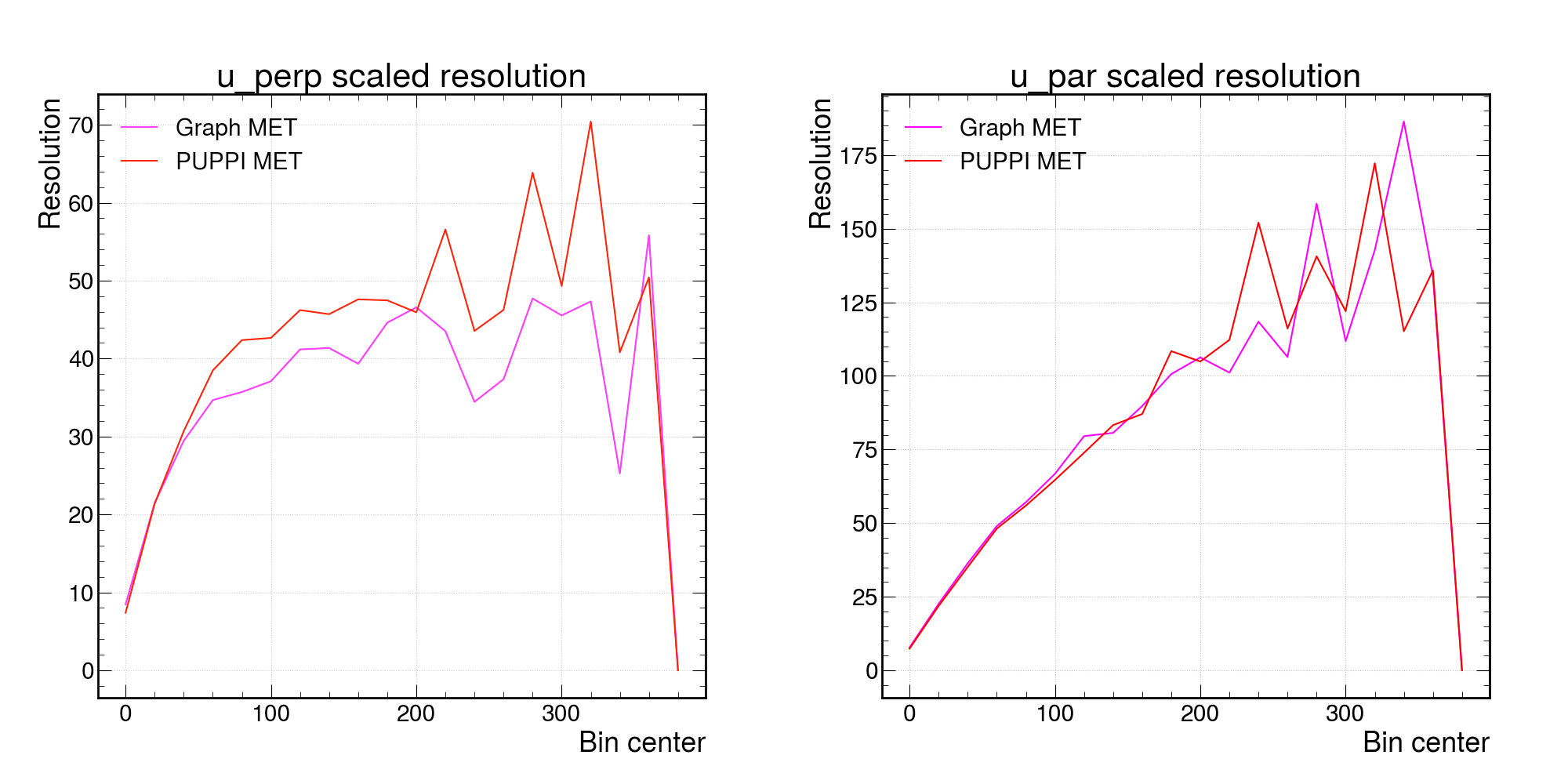}}
\caption{MET resolutions with Dynamic GNN vs. traditional PUPPI algorithm (bin center = bin of MET values where corresponding resolution is computed, lower resolution = higher similarity between true and reconstructed values, perp and par = MET vector's perpendicular and parallel components)}
\label{fig:graph_met}
\end{figure}

We trained a novel EdgeConv-based Dynamic GNN for next-generation L1T system to perform regression of total missing transverse energy (MET) sum (an important quantity that estimates momentum scale and imbalance of particle interactions in collision events but suffers from accumulating effects of detector noise, acceptance gaps, etc). Our new GNN solution, the \textit{L1DeepMETv2} model, utilized graph approach (with nodes as particles and edges as their relationships) and exploited interdependencies between particles reconstructed by L1T. As in Fig.~\ref{fig:graph_met}, our approach allowed \textbf{MET resolution enhancement} in MET vector's perpendicular component with respect to traditional PUPPI algorithm (which computed fixed, local weights per particle simply based on neighbors, according to~\cite{L1TTDR}). With ultra-low-latency goal of 12.5~µs for deployment by 2030, this demands iterative research to achieve the desired dynamic GNN inference.

\begin{figure}[!t]
\centerline{\includegraphics[width=\columnwidth]
{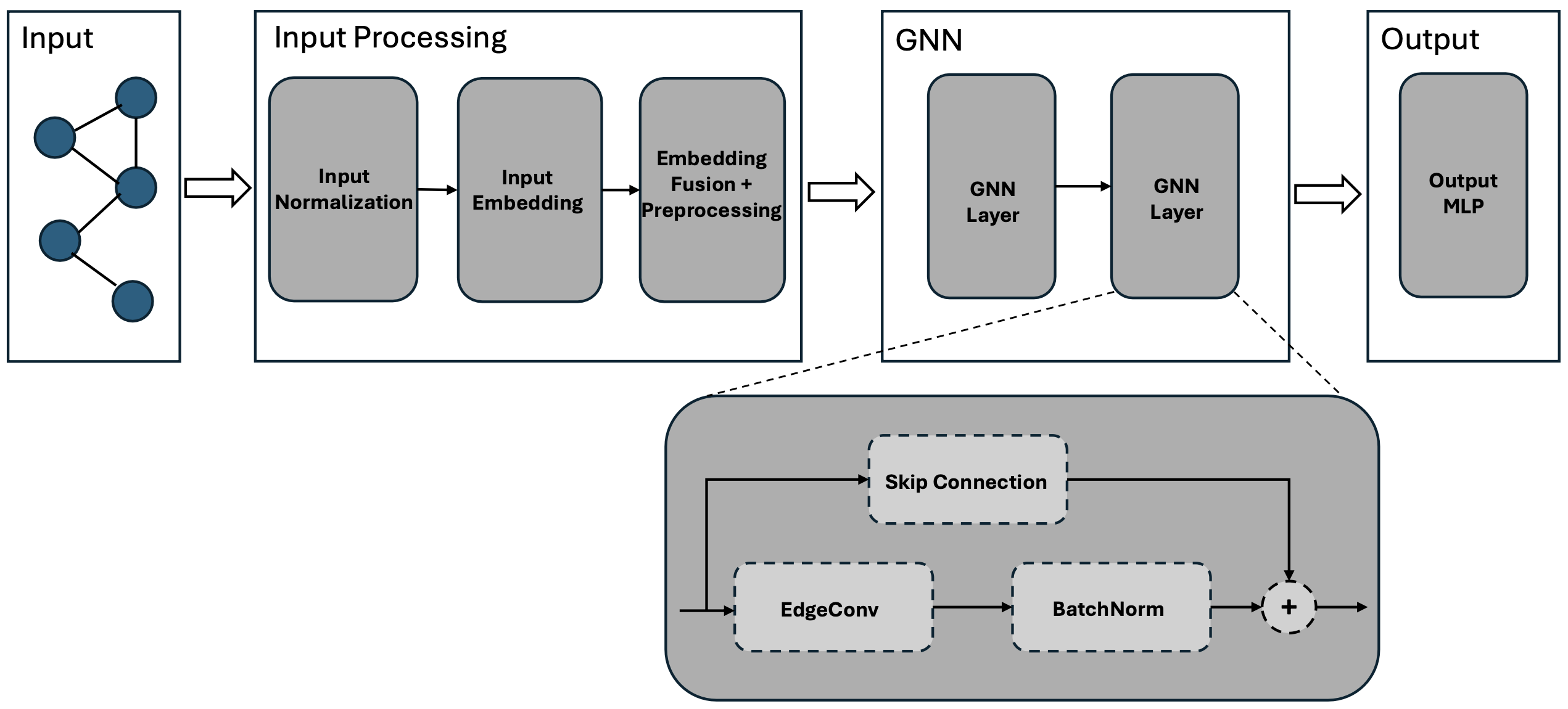}}
\caption{The L1DeepMETv2 model architecture}
\label{fig:L1DeepMETv2_arch}
\end{figure}

Model evaluation consists of three main stages, as illustrated in Fig.~\ref{fig:L1DeepMETv2_arch}. The first stage consists of normalizing and embedding per-particle input embeddings. Continuous and categorical embeddings are concatenated, followed by MLP and batch normalization to provide the initial node embeddings. The second stage passes these embeddings through two message-passing GNN layers. Each message-passing GNN layer includes an EdgeConv~\cite{wang2019dynamicgraphcnnlearning} that computes messages from particle pairs' embeddings and aggregates the neighbor information, a batch normalization, and a residual connection for stable performance optimization. The final stage includes MLP that projects final node embeddings to per-particle weight for regression of reconstructed MET sum.

\subsection{Dynamic Graph Construction}
The stochastic nature of collision events implies that per-event graphs are dynamically constructed with edges created between nodes $u$ and $v$ if particles are within spatial proximity:
\begin{equation}
    \Delta R^2(u,v) = (\eta_u-\eta_v)^2 + (\phi_u-\phi_v)^2<\delta^2
    \label{eqn:dR}
\end{equation}
where $\eta$ and $\phi$ are pseudorapidity and azimuthal angle in CMS coordinate system and $\delta$ is tunable distance threshold.

\begin{figure}[!b]
\centerline{\includegraphics[width=\columnwidth]
{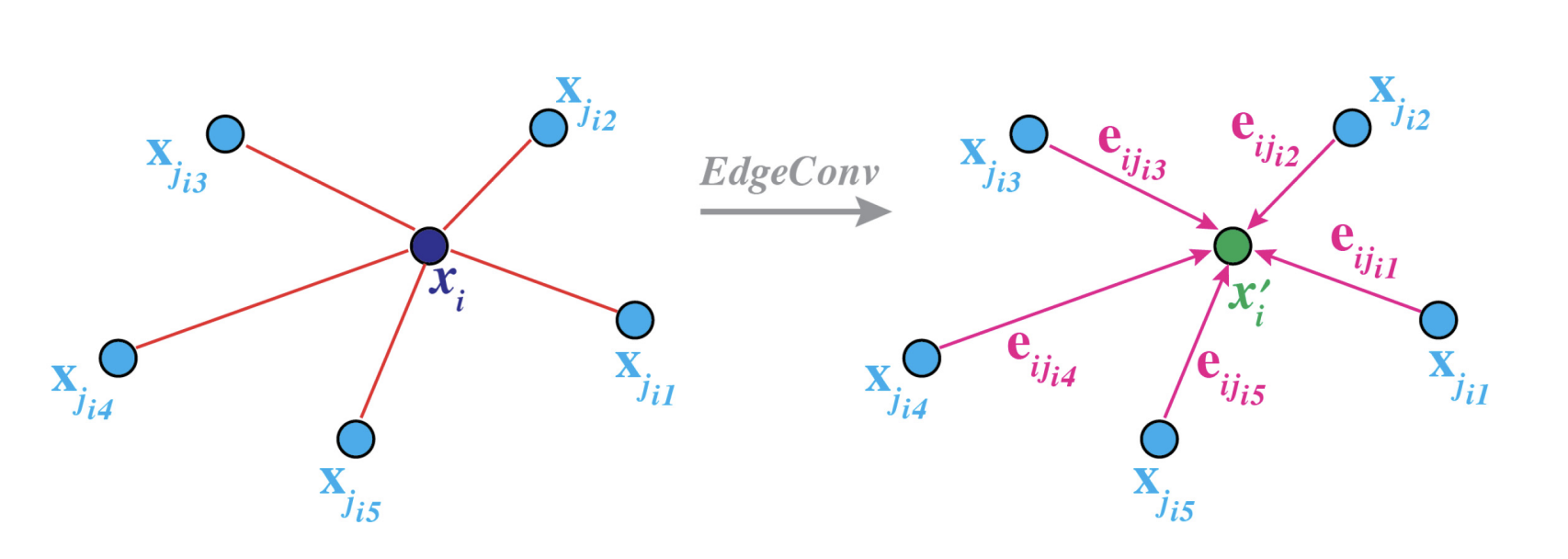}}
\caption{The EdgeConv operation, adapted from \cite{wang2019dynamicgraphcnnlearning}.}
\label{fig:edgeconv}
\end{figure}

\subsection{EdgeConv Operator}
EdgeConv~\cite{wang2019dynamicgraphcnnlearning}, as shown in Fig.~\ref{fig:edgeconv}, is the core component of L1DeepMETv2 model architecture. It computes the messages between connected nodes based on the embeddings of both source node $u$ and neighbor target node $v$ as:
\begin{equation}
m_{uv} = e_{uv} = \phi(x_u, x_v - x_u)
\label{eqn:muv}
\end{equation}
where $\phi$ is learnable function, particularly lightweight MLP. ($x_v - x_u$) enables encoding local relationships between neighbor nodes to represent correlations among particles. This requires graph's dynamic in-place updates of edge embeddings with availability of both source and target node embeddings.

\section{Baseline Hardware Architecture - FlowGNN}
FlowGNN~\cite{FlowGNN} is a modular, dataflow architecture to accelerate various message-passing GNNs on FPGA. It utilizes $P_{edge}$ number of Message Passing (MP) Units that compute edge embeddings in parallel, $P_{node}$ number of Node Transformation (NT) Units that apply transformations for node embeddings in parallel, and an adapter that connects to MP Units and NT Units via streaming FIFOs (First-In First-Out) and multicasts data to the correct processing units. The architecture also employs memory usage optimizations. It uses double data buffers that switch roles of reading and writing data across GNN layers. The architecture also supports storing graph data in compressed sparse row (CSR) format for efficient storage of sparse and irregular graphs.

However, FlowGNN focuses on inference across static graph topologies, where edge lists and embeddings are fixed. Therefore, it lacks support for runtime edge embedding computation required by edge-based dynamic GNN inference.

\section{Proposed Hardware Architecture - DGNNFlow}
We propose a novel streaming dataflow hardware architecture DGNNFlow, which extends FlowGNN \cite{FlowGNN} to enable real-time edge-based dynamic GNN inference applications (where edge embeddings can dynamically change based on neighbor node embeddings). DGNNFlow resolves FlowGNN's limitation in inference with only static graph topologies and lack of support for dynamic edge embedding computation.

\subsection{Overall Dataflow Execution}
\begin{figure*}[t]
\centerline{\includegraphics[width=2\columnwidth]
{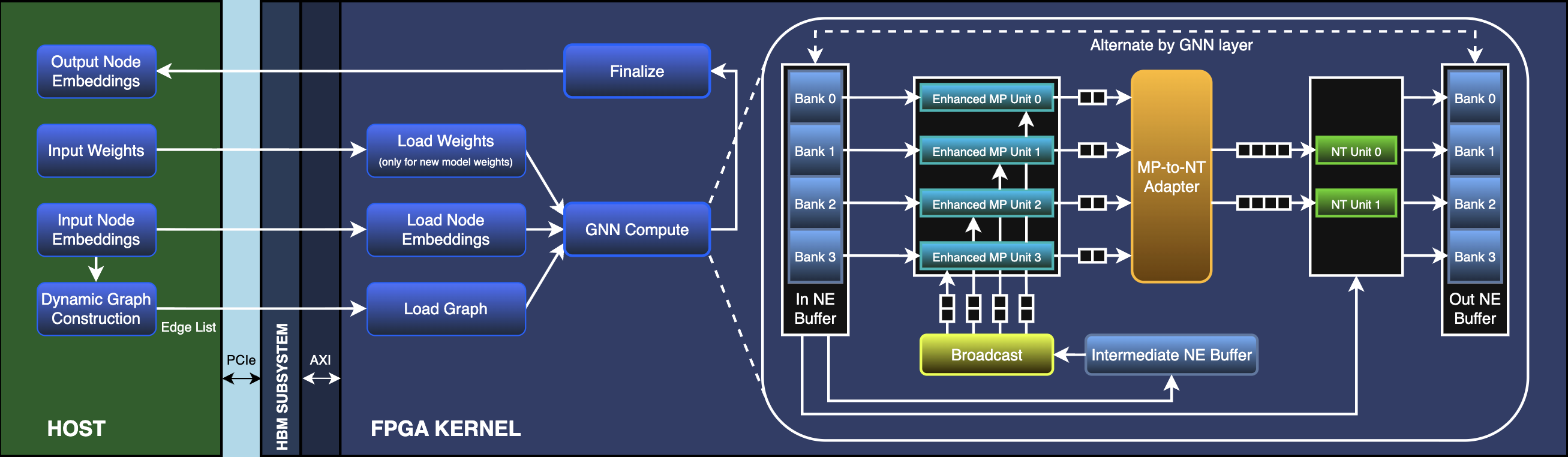}}
\caption{DGNNFlow Architecture for Real-time Edge-based Dynamic GNN Inference}
\label{fig:dataflow}
\end{figure*}

Our proposed architecture is illustrated in Fig.~\ref{fig:dataflow}. Dataflow for each GNN layer starts from BRAM-based Input Node Embedding (NE) buffer, which is partitioned into $P_{edge}$ banks and can be read by $P_{edge}$ Enhanced MP Units in parallel to receive source node embeddings. Node Embedding Broadcast stores copy of node embeddings from Input NE buffer in BRAM-based Intermediate NE buffer and broadcasts target node embeddings via streaming FIFOs to all Enhanced MP Units. Each Enhanced MP Unit selectively captures target node embeddings, computes edge embeddings and sends messages via streaming FIFOs to MP-to-NT adapter, which aggregates messages and sends to the corresponding NT Units via streaming FIFOs. Each NT Unit uses messages received and copy of Input NE buffer to compute the resulting node embeddings of the current GNN layer. $P_{node}$ MP Units simultaneously write the resulting node embeddings to $P_{node}$ banks among total $P_{edge}$ banks of BRAM-based Output NE buffer at a time. After write completion for Output NE buffer, Input and Output NE buffers are swapped for the subsequent GNN layer.

The subsequent three sections explain the novel functionalities we introduce in DGNNFlow: Enhanced MP Units, Node Embedding Broadcast, and Input Dynamic Graph Construction. These functionalities enable complete support of real-time edge-based dynamic GNN inference with dynamic edge embedding computations directly on FPGA.

\subsection{Enhanced MP Units}
Unlike conventional GNNs that rely on static, pre-computed edge embeddings, edge-based dynamic GNNs like EdgeConv~\cite{wang2019dynamicgraphcnnlearning} infer edge embeddings at runtime based on embeddings of both source and target nodes.

In FlowGNN~\cite{FlowGNN}, each MP Unit can locally access its assigned source node embedding, but has \textbf{no local memory access to target node embeddings}. Without this access, it is impossible to compute edge embeddings in edge-based dynamic GNNs without significant communication overhead or memory duplication, incurring FlowGNN's shortcoming in computing edge embedding at runtime. We designed Enhanced MP Units to dynamically compute edge embeddings.

\begin{algorithm}
\caption{Enhanced MP Unit}
\begin{algorithmic}[1]
\STATE Receive target node embeddings from Node Embedding Broadcast via streaming FIFOs.
\STATE Select target node embeddings $x_v$ where degree of $v$ within this Enhanced MP Unit $\neq 0$.
\FOR{each target node embedding $x_v$ selected}
    \IF{target node $v$ matches assigned edge $(u, v)$}
        \STATE Retrieve node embedding $x_u$ of source node $u$.
        \STATE Compute difference vector $\triangle \mathbf{x} = x_v - x_u$.
        \STATE Concatenate $(x_u, \triangle \mathbf{x})$ and pass the concatenated embedding through message function $\phi$ to produce edge message $\mathbf{m}_{uv}$.
        \STATE Stream edge message $\mathbf{m}_{uv}$ to MP-to-NT adapter via streaming FIFOs.
    \ENDIF
\ENDFOR
\end{algorithmic}
\label{algo:edgeembedding}
\end{algorithm}

Overall operation of Enhanced MP Unit is illustrated in Algorithm~\ref{algo:edgeembedding}. Enhanced MP Unit is structured into three key tasks as described in the following three subsections. We perform dataflow pipelining throughout all three tasks (with FIFOs between consecutive tasks) to enable overlapping tasks' operations for task-level parallelism.

\subsubsection{Degree Reading and Target Node Embedding Selection} Enhanced MP Unit listens to broadcast target node embeddings. For each node, if the node's degree within the current Enhanced MP Unit is non-zero, target node embedding is buffered for message computation.

\subsubsection{Message Gathering (EdgeConv Computation)} For each matching source node, Enhanced MP Unit computes difference between source node embedding and buffered target node embedding, concatenates source node embedding and the difference into single embedding vector, applies linear transformation (matrix multiplication followed by bias addition), and uses max-pooling aggregation across transformed neighbor embeddings to obtain the final message. We utilize loop unrolling in vectorized operations (embedding concatenation, matrix multiplication, max-pooling) for SIMD-based parallelism across embedding dimensions.

\subsubsection{Message Expansion} Results from partial message aggregation are expanded to the full set of destination nodes and streamed to MP-to-NT adapter.

\subsection{Node Embedding Broadcast}
To alleviate data dependencies in edge-based dynamic GNN dataflow, we introduce Node Embedding Broadcast streaming module with overall workflow as described in Algorithm~\ref{algo:nodeembedding}. This broadcasting approach fits naturally into streaming hardware model since it eliminates the need for irregular memory access to embedding matrix for target node embeddings.

\begin{algorithm}
\caption{Node Embedding Broadcast}
\begin{algorithmic}[1]
\STATE Retrieve the complete node embedding matrix $\{x_v\}$ from BRAM-based Intermediate Node Embedding Buffer.
\FOR{each target node embedding $x_v$}
    \STATE Broadcast target node embedding $x_v$ to all Enhanced MP Units via streaming FIFOs.
    \FOR{each Enhanced MP Unit}
        \STATE Receive and select node embeddings $x_v$ for message computation as described in Algorithm 1.
    \ENDFOR
\ENDFOR
\end{algorithmic}
\label{algo:nodeembedding}
\end{algorithm}

Node Embedding Broadcast streaming module broadcasts each node embedding simultaneously to parallel FIFOs, where Enhanced MP Unit can receive its own copy of the node embeddings and independently select only target node embeddings with matching source nodes. Specifically, we perform dataflow pipelining throughout broadcast and MP units for concurrent tasks' execution and loop unrolling on the embedding broadcast loop for simultaneous streaming into multiple independent FIFOs, preventing serialization bottlenecks.

To fully reinforce our approach, we also consider possible alternative designs with their functionalities and trade-offs:
\begin{itemize}
\item Full Replication: Each MP Unit locally stores the entire node embedding matrix. This approach requires multiple duplications of node embedding matrix, high memory usage, and poor scalability for large graphs.
\item Selectively Multicast Bus: A specialized bus selectively pushes destination node embeddings to MP Units. This approach incurs complex control, routing congestion, and scalability bottleneck.
\end{itemize}

Node Embedding Broadcast supports three crucial features. First, by broadcasting all node embeddings and offloading dynamic filtering of target nodes to multiple Enhanced MP Units, it prevents issues of alternative approaches, including complex control in single multicast bus (in Selectively Multicast Bus) or redundant local storage of entire node embedding matrix per MP Unit (in Full Replication), and allows for more efficient workload balancing between Node Embedding Broadcast and Enhanced MP Units. Second, all node embeddings can be transmitted across Enhanced MP Units in deterministic manner. Third, the design naturally accommodates graphs with varying degrees of sparsity and maintains efficiency with growth in graph size.

\subsection{Input Dynamic Graph Construction}
We perform input dynamic graph construction in the host side. Host processor constructs input edge list based on Euclidean distance thresholding as in Equation~\ref{eqn:dR} to form event-specific graph connectivity. The resulting list of edges and the node embedding matrix are packed into memory buffers, sent over PCIe to HBM subsystem, and accessed as inputs by on-chip edge-based dynamic GNN inference kernel via AXI interface. The edge list is consumed by the Graph Compute stage to construct internal representations (degree and neighbor tables) for integration with message passing operations in edge-based dynamic GNN inference.

\section{Experiments and Evaluation}
\subsection{Implementation and Model Configurations}
We implemented DGNNFlow architecture for real-time edge-based dynamic GNN inference using High-Level Synthesis with Vitis, Vivado and orchestrated data communication between host and FPGA using OpenCL. We cross-checked our implementation with PyTorch implementation of L1DeepMETv2 to ensure consistent end-to-end execution. We deployed DGNNFlow using AMD Alveo U50 FPGA board for performance evaluation with 200~MHz clock frequency. Resource utilization of DGNNFlow is reported by Vitis Analyzer and recorded in Table~\ref{table:resource}.

\subsection{Dataset and Baselines}
We used DELPHES framework~\cite{Delphes} to simulate HL-LHC proton-proton collision events, generating test dataset of 16K graphs as direct inputs of L1DeepMETv2. For baselines, we measured performance on CPU (Intel Xeon Gold 6226R) and GPU (NVIDIA RTX A6000) under two software (SW) variants: Baseline SW version as PyTorch model and Optimized SW version, which utilizes \verb|torch.compile| \cite{TorchCompile} with Just-In-Time compilation of PyTorch code for kernel optimization. For GPU, we evaluated batching with sizes from 1 to 8 to understand latency per graph against DGNNFlow. It is important to note that due to HL-LHC trigger system's event-triggered nature (with 1 input graph event at a time) and the real-time constraint, \textbf{only E2E (end-to-end) latency with batch size 1} should be used for final evaluation.

\subsection{End-to-end Evaluation}

\begin{table}[b]
\caption{Resource Availability and Usage on AMD Alveo U50 FPGA}
\begin{center}
\begin{tabular}{|c|c|c|c|c|}
\hline
Resource & \textbf{LUT} & \textbf{Register}& \textbf{BRAM}& \textbf{DSP} \\
\hline
Available & 872K & 1,743K & 1,344 & 5,952 \\
Usage & 235,017 & 228,548 & 488 & 601 \\
\hline
\end{tabular}
\label{tab1}
\end{center}
\label{table:resource}
\end{table}

\begin{figure}[b]
\centerline{\includegraphics[width=\columnwidth]
{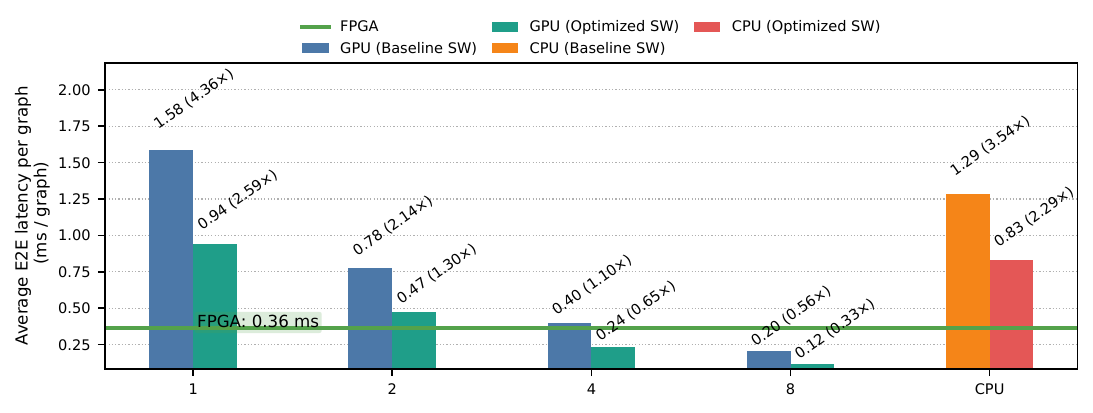}}
\caption{Average E2E Latency per graph by Batch size}
\label{fig:perf_by_batches}
\end{figure}

\begin{figure}[t]
\centerline{\includegraphics[width=\columnwidth]
{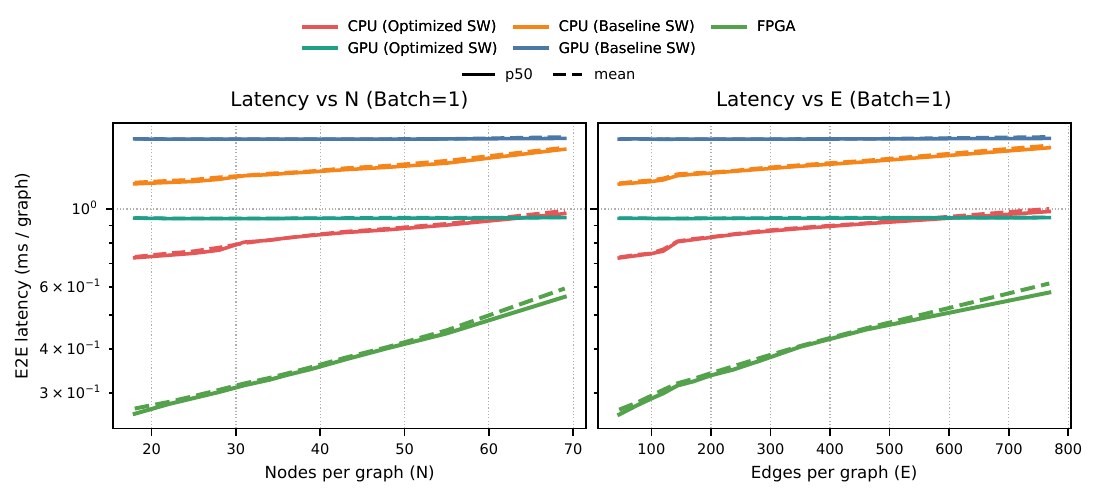}}
\caption{E2E Latency per graph by Numbers of Nodes and Edges}
\label{fig:perf_by_NE}
\end{figure}

We evaluated E2E latency for DGNNFlow, CPU, and GPU. E2E latency consists of data transfer time between host and device (for DGNNFlow, GPU) and edge-based dynamic GNN inference runtime. Input dynamic graph construction time is only for generation of model's input and excluded from E2E latency for all of DGNNFlow, CPU, GPU to allow for measurement of model's actual execution time. Fig.~\ref{fig:perf_by_batches} illustrates average E2E latency per graph for GPU at batch sizes 1 to 8 and for CPU, DGNNFlow at batch size 1. DGNNFlow achieves 0.36~ms per graph. DGNNFlow is 3.54x and 2.29x faster than CPU Baseline SW and Optimized SW versions respectively. DGNNFlow outperforms GPU Baseline SW version by 4.36x, 2.14x and outperforms GPU Optimized SW version by 2.59x, 1.30x for batch sizes 1 and 2 respectively. Higher batch size reduces GPU's average E2E latency per graph, so GPU can deliver higher throughput when batching. However, due to real-time, event-triggered nature of HL-LHC trigger systems, only batch size of 1 remains the main comparison point.

Fig.~\ref{fig:perf_by_NE} shows latency distribution by numbers of nodes and edges at batch size 1. As graph size grows, CPU latency increases steadily for both Baseline SW and Optimized SW versions. GPU latency is high with small graph but stays highly consistent with graph size growth due to better utilization of GPU's high-throughput capability. Despite increasing latency over graph size growth, DGNNFlow still achieves remarkably shorter latency than both CPU and GPU.

We also evaluated average power consumption for FPGA system (with HBM) for DGNNFlow, CPU, and GPU (at batch size 1) as shown in Table~\ref{table:power}. FPGA system's power consumption (from the two power rails) is about 3.59x - 3.70x less than GPU, about 1.93x - 2.12x less than CPU, and internal FPGA power consumes only about 7.684W.

\begin{table}[t]
\caption{Average Power Consumption Among \\ GPU, CPU, and FPGA System (with HBM) for DGNNFlow}
\begin{center}
\begin{tabular}{|c|c|c|c|}
\hline
\textbf{\makecell{GPU\\Baseline SW}} & \textbf{\makecell{GPU\\Optimized SW}} & \textbf{\makecell{CPU\\Baseline SW}} & \textbf{\makecell{CPU\\Optimized SW}} \\
\hline
75.661 W & 77.995 W & 44.650 W & 40.588 W \\
\hline
\multirow{2}{*}{\textbf{\makecell{FPGA System \\ (with HBM)}}}
& \textbf{Internal} & \textbf{3.3V PCIe} & \textbf{12V PCIe} \\ 
\cline{2-4}
& 7.684 W & 1.935 W & 19.145 W \\
\hline
\end{tabular}
\label{tab2}
\end{center}
\label{table:power}
\end{table}

Therefore, DGNNFlow's main strength is low-latency, low-power inference while GPU’s advantage remains throughput with batching. Although the long-term goal of 12.5~µs latency has not been met yet, DGNNFlow provides the first pivotal step for acceleration with novel real-time edge-based dynamic GNN inference approach in HL-LHC trigger systems.

\section{Conclusion}
This paper presented DGNNFlow, a streaming dataflow architecture for real-time edge-based dynamic GNN inference applications, especially HL-LHC trigger systems, with three novel functionalities as extension of FlowGNN~\cite{FlowGNN}, including Enhanced MP Units, Node Embedding Broadcast, and Input Dynamic Graph Construction. DGNNFlow exhibited 2.59x - 4.36x and 1.30x - 2.14x speedup compared to NVIDIA RTX A6000 GPU (at batch sizes 1 and 2) with 3.59x - 3.70x less power consumption, achieved 2.29x - 3.54x speedup with 1.93x - 2.12x less power consumption compared to Intel Xeon Gold 6226R CPU. DGNNFlow provides the first pivotal step for hardware acceleration of edge-based dynamic GNN inference in HL-LHC trigger systems and is effectively applicable for edge-based dynamic GNN models.

\bibliographystyle{IEEEtran}
\bibliography{references}

\end{document}